\documentclass[english,pra,aps]{revtex4}
\usepackage[T1]{fontenc}
\usepackage[latin1]{inputenc}
\usepackage{graphicx}

\makeatletter

 \usepackage{dcolumn}
 \usepackage{bm}
 \usepackage{subfigure}

\usepackage{babel}
\makeatother

\begin{document}

\title{Analytic treatment of CRIB Quantum Memories for Light using Two-level
Atoms }

\author{J.~J.~Longdell} \affiliation{Department of Physics, University of Otago, Dunedin, New Zealand} 
\email{jevon.longdell@otago.ac.nz}
\author{G.~H\'{e}tet} \affiliation{ARC COE for Quantum-Atom Optics, Australian National   University, Canberra, ACT 0200, Australia}
\author{M. J. Sellars} \affiliation{Laser Physics Centre, RSPhysSE, Australian National   University, Canberra, ACT 0200, Australia}
\author{P. K. Lam} \affiliation{ARC COE for Quantum-Atom Optics, Australian National   University, Canberra, ACT 0200, Australia}

\date{1 July 2008}

\begin{abstract}
It has recently been discovered that the optical analogue of a gradient
echo in an optically thick material could form the basis of a optical
memory that is both completely efficient and noise free. Here we present
analytical calculation showing this is the case. There is close analogy
between the operation of the memory and an optical system with two
beam splitters. We can use this analogy to calculate efficiencies
as a function of optical depth for a number of quantum memory schemes
based on controlled inhomogeneous broadening. In particular we show
that multiple switching leads to a net 100\% retrieval efficiency
for the optical gradient echo even in the optically thin case.
\end{abstract}



\maketitle

\section{Introduction}

The ability to store and recall quantum states of light as coherences
in atomic media is currently being actively pursued. Such quantum
memories would find use both in optical based quantum computation
\cite{knil01} and long distance quantum communication \cite{simo07}. 

Many of the current quantum memory approaches involve the use of three
level atomic systems, and store quantum information between two quasi
ground states \cite{juls04,flei02,gors07,mois01,krau06}. Quantum
states have been stored and recalled using this approach \cite{juls04,chan05,eisa05,appe08}.
Three level schemes are particularly well suited to gaseous atomic
systems where significant optical depths can be obtained for the optical
transitions due the allowed transitions and long coherence times can
be obtained for the ground state coherences. Rare earth ion doped
solids at cryogenic temperatures have much higher atom densities and
allow reasonable optical thickness to be obtained from ensembles of
weak oscillators. These weak oscillators can have correspondingly
long coherence times. This means that approaches involving only two
level atoms can be considered. Sangouard et al. \cite{sang07} showed
that in principle 54\% efficiency can be achieved from a controlled
reversible inhomogeneous broadening (CRIB) echo \cite{mois01,krau06}
with two level atoms, in the case where the broadening mechanism is
`transverse', that is not correlated with position along the optical
path. The efficiency was limited by reabsorbtion, as the sample is
made more optically thick, which is required to absorb the input pulse,
more of the echo gets absorbed before it can make it out of the sample.
Shortly afterward it was shown that the optical gradient echo or {}``longitudinal''
CRIB echo \cite{alex06}, the only CRIB echo that has so far been
reported experimentally, did not suffer from reabsorbtion problems
and offered noise free, potentially 100\% efficient storage \cite{hete08}.
An attractive property of these echo based techniques is that time
bandwidth product scales better with optical depth \cite{hete08b,simo07}.
Indeed storing and recalling multiple pulses has been demonstrated
for CRIB echo memories \cite{alex07} but this has proved difficult
for EIT.

Following on from \cite{hete08} here we present analytic theory of
the optical gradient echo or {}``longitudinal CRIB'' echo. Analytic
solutions of the Maxwell-Bloch equations are derived for a pulse of
light interacting with ensemble of atoms, where the resonant frequency
atoms varies linearly with the propagation distance. After the input
pulse atomic coherence dephases due to the different resonant frequencies
of the atomic ensemble. If the detunings of each of the atoms are
reversed the ensemble rephases and produces an echo. In the case where
the medium is optically thick the echo is totally efficient. Indeed,
we show very close analogy between a gradient echo memory and a pair
of beamsplitters. We use this analogy to calculate the efficiency
of the longitudinal echo for optically thin samples. Multiple switching
and a number of memories operated in series can also be easily treated
using this beamsplitter analogy. As transverse broadening can be considered
as the limit of many stacked optically think transverse broadening.
Transverse broadening can be treated using the same approach. We reproduce
previous results for the efficiency of transverse CRIB echoes \cite{sang07}
and extend to the case of multiple switching.

\section{Theory}

We consider the interaction of a collection of two-level atoms with
a light field where a detuning of the atoms that is linearly dependent
on their position can be introduced. We shall assume that the area
of the incoming pulses is much less than that of a $\pi$ pulse. This
enables us to treat the atoms as harmonic oscillators. Before the
detuning of the ions are flipped, the Maxwell-Bloch equations in the
frame at the speed of light are \cite{cris70} \begin{eqnarray}
\frac{\partial}{\partial t}\alpha(z,t) & = & -(\gamma+i\eta z)\alpha(z,t)+igE(z,t)\label{eq:mb1}\\
\frac{\partial}{\partial z}E(z,t) & = & iN\alpha(z,t)\label{eq:mb2}\end{eqnarray}
 Where $E$ represents the slowly varying envelope of the optical
field; $\alpha$ the polarization of the atoms; $N$ the atomic density;
$g$ the atomic transition coupling strength; $\gamma$ the decay
rate from the excited state and $\eta z$ the detuning from resonance.
We shall assume that the process happens fast compared to the atomic
decay rate and will take $\gamma=0$.

After the detuning of the ions is flipped to recall the light pulse
the Maxwell-Bloch equations describing the dynamics are the same as
above except the sign of the $i\eta z$ term in \ref{eq:mb1} is flipped
leading to (with $\gamma=0$)

\begin{eqnarray}
\frac{\partial}{\partial t}\alpha(z,t) & = & -i\eta z\alpha(z,t)+igE(z,t)\label{eq:mb3}\\
\frac{\partial}{\partial z}E(z,t) & = & iN\alpha(z,t)\label{eq:mb4}\end{eqnarray}

As mentioned above, although the treatment here is classical, the
linearity of Eqs.~(\ref{eq:mb1},\ref{eq:mb2},\ref{eq:mb3},\ref{eq:mb4}
) ensures that exactly the same analysis would be true for operator
valued $\alpha$ and $E$. The only added noise in the output pulses
will be the vacuum noise added to preserve commutation relations in
much the way as light interacting with a beam-splitter. Our results
will thus hold for quantum mechanical fields also. 
\begin{figure}
\begin{centering}
\includegraphics[width=0.8\textwidth]{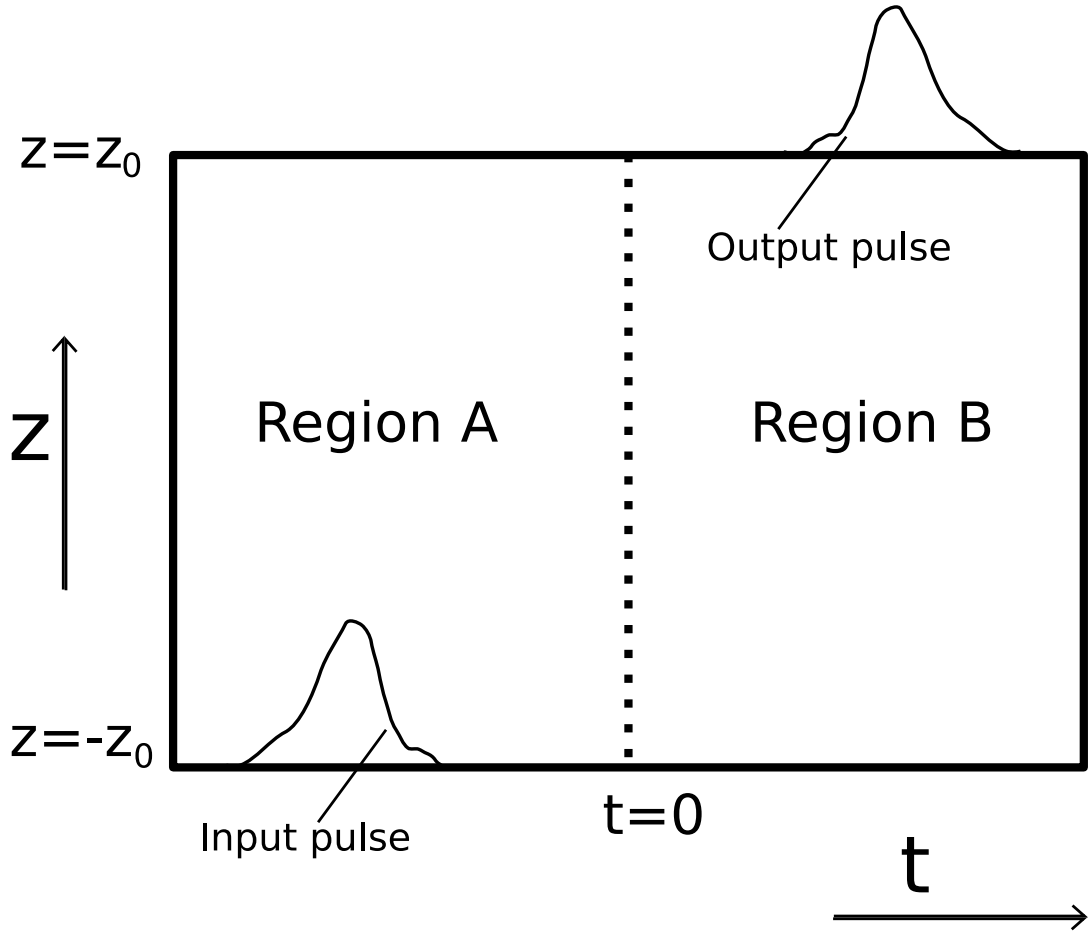} 
\par\end{centering}

\caption{The domain over which the Maxwell-Bloch equations are solved. Each
horizontal slice represents one position in the sample of atoms as
a function of time. The light enters the sample at $z=-z_{0}$ and
leaves at $z=+z_{0}$. The dotted line represents the point in time
at which the detunings of the ions are flipped. In Region A the dynamics
are described by Eqns. \ref{eq:mb1} and \ref{eq:mb2}, in Region
B the dynamics are described by Eqns. \ref{eq:mb3} and \ref{eq:mb4}.
\label{fig:domain} }

\end{figure}

The domain over which these equations will be solved is illustrated
in Fig.~\ref{fig:domain}. The boundary conditions are the input
pulse $E(z=-z_{0},t)=f_{\textrm{in}}(t)$ and the initial state of
the atoms. We shall assume that the atoms start in their ground state
so that $\alpha(z,t=-\infty)=0$. By propagating forward in both $z$
and $t$ from these boundary conditions one could determine the final
state of the atoms and the light exciting the sample at $z=z_{0}$.
Ideally the procedure would result in an output pulse, $E(z=+z_{0},t)=f_{\textrm{out}}(t)$
that has the same energy and is closely related to the input pulse,
and that at the end of the process all the atoms are left in their
ground states, $\alpha(z,t=+\infty)=0$. It should be noted that we
are dealing with the situation where there is no decay of the atomic
states, so in order for $\alpha(z,t=+\infty)=0$ the atoms must be
left in their ground states at the end of the retrieval process, it
cannot happen via decay of the atomic states.

Rather than propagating forward all the way from our boundary conditions
at $z=-z_{0}$ and $t=-\infty$ to arrive at expressions for $z=z_{0}$
and $t=+\infty$ we will take a different approach that better makes
use of the symmetry of the situation. We first solve for the behavior
in Region A using the boundary conditions $\alpha(z,t=-\infty)=0$
and $E(z=-z_{0},t<0)=f_{\textrm{in}-}(t)$. This analysis will give
us expressions for $E$ and $\alpha$ at the time when the detunings
are flipped ($\alpha(z,t=0)$ and $E(z,t=0)$) as well an expression
for the light that leaves the sample before the field is flipped ($E(z=+z_{0},t<0)=f_{\textrm{out}-}(t)$).

We then carry out what turns out to be a very similar analysis. Working
in Region B, we start with an arbitrary form output pulse ($E(z=+z_{0},t>0)=f_{\textrm{out}+}(t)$)
and the requirement that the atoms end up in their ground states ($\alpha(z,t=\infty)=0$)
and then we work out what is needed from the other two boundary conditions
E($z=-z_{0},t>0$) and ($\alpha(z,t=0)$ and $E(z,t=0)$) in order
for these outcomes to occur. By comparing these required boundary
conditions with those that actually occur we show that in the case
of an optically thick sample the storage is completely efficient.
For the case of an optically thin sample we can calculate how efficient
the process will be.

The analysis of the behavior in Region A is as follows: First Eq.~\ref{eq:mb1}
is integrated to give

\begin{eqnarray}
\alpha_{a}(z,t) & = & ig\int_{-\infty}^{t}dt'\, e^{-i\eta z(t-t')}E_{a}(z,t')\nonumber \\
 & = & ig\int_{-\infty}^{\infty}dt'\, H(t-t')e^{-i\eta z(t-t')}E_{a}(z,t')\end{eqnarray}
 here $H$ denotes the Heaviside step function. The above expression
is in the form of a convolution, taking the Fourier transform gives
a product.

\begin{equation}
\alpha_{a}(z,\omega)=igE_{a}(z,\omega)\left(\frac{1}{i(\omega+\eta z)}+\pi\delta(\omega+\eta z)\right)\end{equation}

Substituting this in Eq.~\ref{eq:mb2} we get

\begin{equation}
\partial_{z}E_{a}(z,\omega)=-g^{2}N\left(\frac{1}{i(\omega+\eta z)}+\pi\delta(\omega+\eta z)\right)E_{a}(z,\omega)\end{equation}
 Integrating this equation we have \begin{eqnarray}
E_{a}(z,\omega) & = & E_{a}(z=-z_{0},\omega)\exp\int_{-z_{0}}^{z}dz'\,-g^{2}N\left(\frac{1}{i(\omega+\eta z')}+\pi\delta(\omega+\eta z')\right)\nonumber \\
 & = & F_{\textrm{in}-}(\omega)\exp(-\beta\pi(H(\omega+\eta z)-H(\omega+\eta z_{0})))\left|\frac{\omega+\eta z}{\omega+\eta z_{0}}\right|^{i\beta}\end{eqnarray}
 where $\beta=\frac{g^{2}N}{\eta}$.

It can be seen that the amplitude of each spectral component is attenuated
by a factor $\exp(-\pi\beta)$ after traveling past the position in
the sample where it is resonant with the atoms. It also receives a
phase shift as it travels through the sample.

We make the assumption that the spectral coverage of the sample is
large compared to the optical depth. That is for each frequency of
interest $\omega$ in our input signal we have $\beta\omega\ll\eta z_{0}$
in which case our expression for $E_{a}(z,\omega)$ takes the form
\begin{equation}
E_{a}(z,\omega)=F_{\textrm{in}-}(\omega)\exp(-\beta\pi H(\omega+\eta z))\left|\frac{\omega+\eta z}{\eta z_{0}}\right|^{i\beta}\label{eq:answer}\end{equation}

Substituting $z=+z_{0}$ in the above equation one finds that the
transmitted pulse is equal to the the incident pulse multiplied by
an attenuation factor $\exp(-\beta\pi)$. \begin{equation}
f_{\textrm{out}-}(t)=f_{\textrm{in}-}(t)e^{-\beta\pi}\label{eq:transmitted_light}\end{equation}

In the limit of large $\beta$ no light is transmitted and all remains
in the material during the period $t<0$.

Integrating Eq.~\ref{eq:answer} with respect to $\omega$ gives
an expression for $E_{a}(z,t=0)$ in the form of a convolution Fourier
transforming along the spatial coordinate we get. \begin{eqnarray}
E_{a}(k,t=0) & = & -f_{\textrm{in}-}(-\frac{k}{\eta})\textrm{sgn}(k)\beta\left|\frac{k}{\eta}\right|^{-1+i\beta}\Gamma(i\beta)\nonumber \\
 &  & \left(\left|\frac{k}{\eta}\right|\cosh\left(\frac{\pi\beta}{2}\right)+\frac{k}{\eta}\sinh\left(\frac{\pi\beta}{2}\right)\right)\label{eq:kspace}\end{eqnarray}
 Here $f_{{\rm {in}}}(k/\eta)$ is the input field at the time $\tau=k/\eta$
and $\Gamma(\xi)$ is the gamma function. An expression for $\alpha$
at the time the field was flipped can easily be obtained from the
above result along with Eq.~\ref{eq:mb2}.

We now turn our attention to Region B and calculate $\alpha_{b}$
and $E_{b}$ in that region subject to our desired boundary conditions
$\alpha_{b}(z,t=\infty)=0$, $E_{b}(z=z_{0},t)=f_{\textrm{out}+}(t)$.
The output field $f_{\textrm{out}+}(t$) is at this stage undetermined. 

Solving Eq.~\ref{eq:mb3} subject to the output condition $\alpha_{b}(z,t=\infty)=0$
gives \begin{eqnarray}
\alpha_{b}(z,t) & = & ig\int_{+\infty}^{t}dt'\, e^{i\eta z(t-t')}E_{b}(z,t')\nonumber \\
 & = & -ig\int_{-\infty}^{\infty}dt'\, H(t'-t)e^{-i\eta z(t'-t)}E_{b}(z,t')\end{eqnarray}
 Fourier transforming gives \begin{equation}
\alpha_{b}(z,\omega)=-igE_{b}(z,\omega)\left(\frac{i}{(\omega-\eta z)}+\pi\delta(\omega-\eta z)\right)\end{equation}

Substituting this in Eq.~\ref{eq:mb4} and integrating gives

\begin{eqnarray}
E_{b}(z,\omega) & = & E_{b}(z=z_{0},\omega)\exp\int_{+z_{0}}^{z}dz'\, g^{2}N\left(\frac{i}{(\omega-\eta z')}+\pi\delta(\omega-\eta z')\right)\nonumber \\
 & = & F_{\textrm{out}+}(\omega)\exp(-\beta\pi(H(\omega-\eta z)-H(\omega-\eta z_{0})))\left|\frac{\omega-\eta z}{\omega-\eta z_{0}}\right|^{i\beta}\end{eqnarray}
 In a similar manner to in Region A, in the limit of large $z_{0}$
this can be approximated. \begin{equation}
E_{b}(z,\omega)=F_{\textrm{out}+}(\omega)\exp(-\beta\pi H(\omega-\eta z))\left|\frac{\omega-\eta z}{\eta z_{0}}\right|^{i\beta}\label{eq:answerb}\end{equation}

From the above expression $E_{b}(z=-z0,t)$ can be calculated resulting
in \begin{equation}
f_{\textrm{in}+}(t)=f_{\textrm{out}+}(t)e^{-\beta\pi}\label{eq:reqinput}\end{equation}

Also from Eq.~\ref{eq:answerb} one can find and expression for $E_{b}(k,t=0)$:

\begin{eqnarray}
E(k,t=0) & = & f_{\textrm{out}+}(-\frac{k}{\eta})\textrm{sgn}(k)\beta\left|\frac{k}{\eta}\right|^{-1-i\beta}\Gamma(-i\beta)\nonumber \\
 &  & \left(\left|\frac{k}{\eta}\right|\cosh\left(\frac{\pi\beta}{2}\right)+\frac{k}{\eta}\sinh\left(\frac{\pi\beta}{2}\right)\right)\label{eq:kspaceb}\end{eqnarray}

Equations~\ref{eq:reqinput} and~\ref{eq:kspaceb} give conditions,
that if satisfied, will lead to a particular output pulse $f_{\textrm{out}+}(t)$
and all the atoms being left in the ground state at the end of the
process. From Eq.~\ref{eq:reqinput} and comparing Eqns.~\ref{eq:kspace}
and~\ref{eq:kspaceb} we see that if an auxiliary input pulse is
applied an output pulse related to our input pulse by\begin{equation}
f_{{\rm {out}}}^{+}(t)=-f_{{\rm {in}}}^{-}(-t)|t|^{2i\beta}\frac{\Gamma(i\beta)}{\Gamma(-i\beta)}.\label{eq:tada}\end{equation}

Because $|t|^{2i\beta}{\Gamma(i\beta)}/{\Gamma(-i\beta)}$ has a modulus
of 1 for all $t$, it can be seen that the envelope of the output
pulse is a time reversed version of the input. The required auxiliary
pulse is given by \ref{eq:kspaceb}, the practical usefulness of the
memory would of course be greatly hampered by the need for this auxiliary
input pulse. Which must be an attenuated copy of the output pulse
applied at the same time as the output pulse. However in the situation
where the sample is optically thick, $\exp(-\beta\pi)\ll1$, the required
input pulse is zero. When arriving at Eq.~\ref{eq:tada} one only
needs to ensure that for $t=0$ the values for $E$ match, the values
for $\alpha$ will then also match because of Eqns.~\ref{eq:mb2}
and~\ref{eq:mb4}.

The phase between the input pulse and the output pulse changes across
the pulse an amount given by $2\beta\log(t_{\text{end}}/t_{\text{start}})$
where $t_{\text{start}}$ and $t_{\text{end}}$ are the start and
end times times for the output pulse. This will be a modest phase
shift in most situations where the memory might be used. A value for
$\beta$ of 2 would provide sufficient optical depth for 99.999\%
efficiency, with this and 2 for $t_{\text{end}}/t_{\text{start}}$
the phase shift across the pulse would be less than $\pi$. This phase
shift could be corrected for by some time dependent change in the
optical path length, for instance a mirror mounted on a piezo or and
electro optic modulator. Alternately two memories could be used in
series with the initial frequency gradients opposite for each memory,
the phase shifts of the two memories would then cancel.

\section{Efficiency and Analogue beamsplitters}

\begin{figure}
\begin{centering}
\includegraphics[width=1\columnwidth]{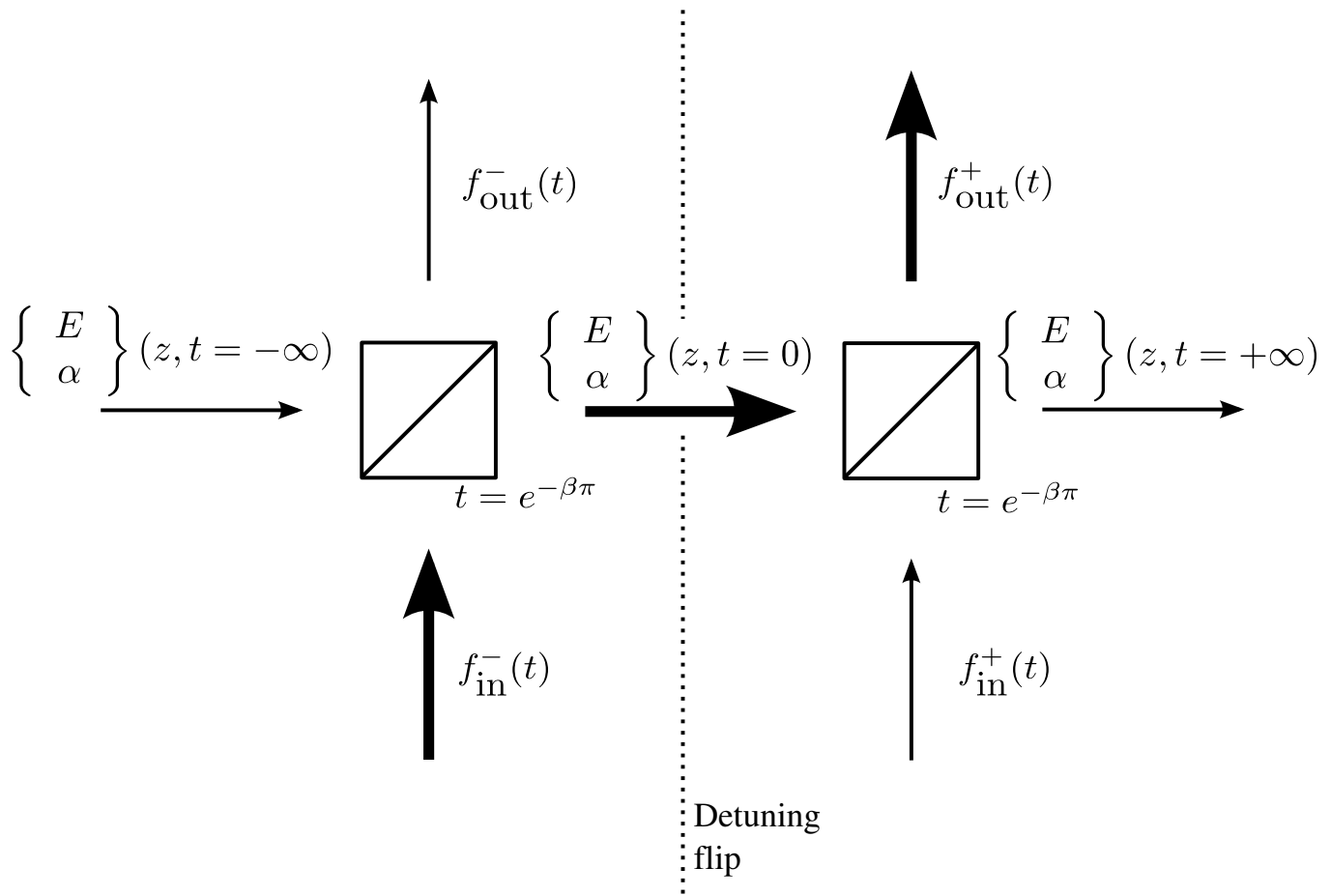}
\par\end{centering}

\caption{The optical gradient echo memory represented as a pair of beam splitters.
The bold arrows represent the flow of energy in the optically thick
case. The spatial and temporal modes that correspond to the each port
of the beam splitter is discussed in the text. \label{fig:twobeamsplitters}}

\end{figure}

The fact that we can still get the desired output pulse with an optically
thin sample by applying and auxiliary input pulse may not be all that
helpful in the practical operation of the memory. It does however,
along with the linearity of the differential equations, enable one
to determine the effect of the memory in the optically thin regime.
The difference between the true output pulse when working in the optically
thin regime and the ideal case is equal to the effect of applying
the auxiliary pulse alone. There is a close analogy between the operation
of the memory in the finite optical depth regime and a beam splitter.
An optical beam splitter takes a pairs of optical modes and outputs
linear combinations of these modes. The action of the memory for both
the times $t<0$ and $t>0$ can be reduced to that of a beam splitter,
these are illustrated in Fig.~\ref{fig:twobeamsplitters} . Each
of the two beamsplitters have two input and two output ports. For
the left hand beamsplitter bottom input port, labeled $f_{\text{}{in}}^{-}(t)$
and the mode of interest is the temporal mode of the optical wavepacket
that we wish to store. The top, output port of the left hand beam
splitter, labeled $f_{\text{}{out}}^{-}(t)$ for this represents the
light that is transmitted through the sample, from Eq.~\ref{eq:transmitted_light},
we can see that this transmitted light has the same temporal mode
as the input light and that the amplitude transmitivity of our analogue
beam splitter will be $\exp(-\beta\pi)$. The left hand input port
and the right hand output ports of the two beamsplitters aren't optical
modes but instead the beamsplitter acts on, and produces, `polariton'
excitations that have both an atomic and optical component. Like the
polaritons considered in electromagnetically induced transparency
(EIT) \cite{flei00} these are combinations of photon and atom excitations.
Unlike EIT polaritons optical gradient polaritons propagate in reciprocal
space, to higher spatial frequencies, rather than propagating along
the propagation direction \cite{hete08b}.

The left hand port of the left hand beamsplitter represents the initial
state of the optical field in the sample and the atoms before the
input pulse is applied in for the case of our memory, during the operation
of the memory this state will be a vacuum state. The right hand output
mode of the first beamsplitter represents the mode of the excitation
in the sample at the time the field is switched $(t=0)$. The optical
component of mode function is given by Eq.~\ref{eq:kspace} and the
atomic part can be found from Eq.~\ref{eq:mb2}. One explanation
for why the memory works is that output polariton mode for the left
hand beamsplitter that represents the $(t<0)$ evolution matches the
input polariton mode for righthand beamsplitter that represents the
(t>0) evolution. So long as we choose input and output optical modes
for the given by equations Eq.~\ref{eq:reqinput} and~\ref{eq:tada}.
The amplitude transmitivity of this second beamsplitter is  given
by $\exp(-\beta\pi)$. 

In order to have an efficient memory we would require excitation fed
into the $f_{\text{in}}^{-}(t)$ port of the network shown in Fig.~\ref{fig:twobeamsplitters}
to be directed entirely to the $f_{\text{out}}^{+}(t)$ port . This
requires high reflectivity for our analogue beamsplitters or equivalently
large optical depth. The beamsplitter analogy easily enables calculation
of the efficiency of the echo process as a function of optical depth.
Because both beamsplitters have an amplitude transmitivity of $\exp(-\beta\pi)$
we end up efficiency for the echo given by\[
\text{Efficiency}=(1-\exp(-2\beta\pi))^{2}\]

This result agrees with the numerical simulations we have presented
previously \cite{hete08}. From the beamsplitter analogue one can
also see immediately where the rest of the incident energy goes, $\exp(-2\beta\pi)$
was transmitted by the sample and $\exp(-2\beta\pi)(1-\exp(-2\beta\pi))$
remains in the sample.

\section{Transverse broadening and multiple switching.}

\begin{figure}
\begin{centering}
\includegraphics[width=0.7\columnwidth]{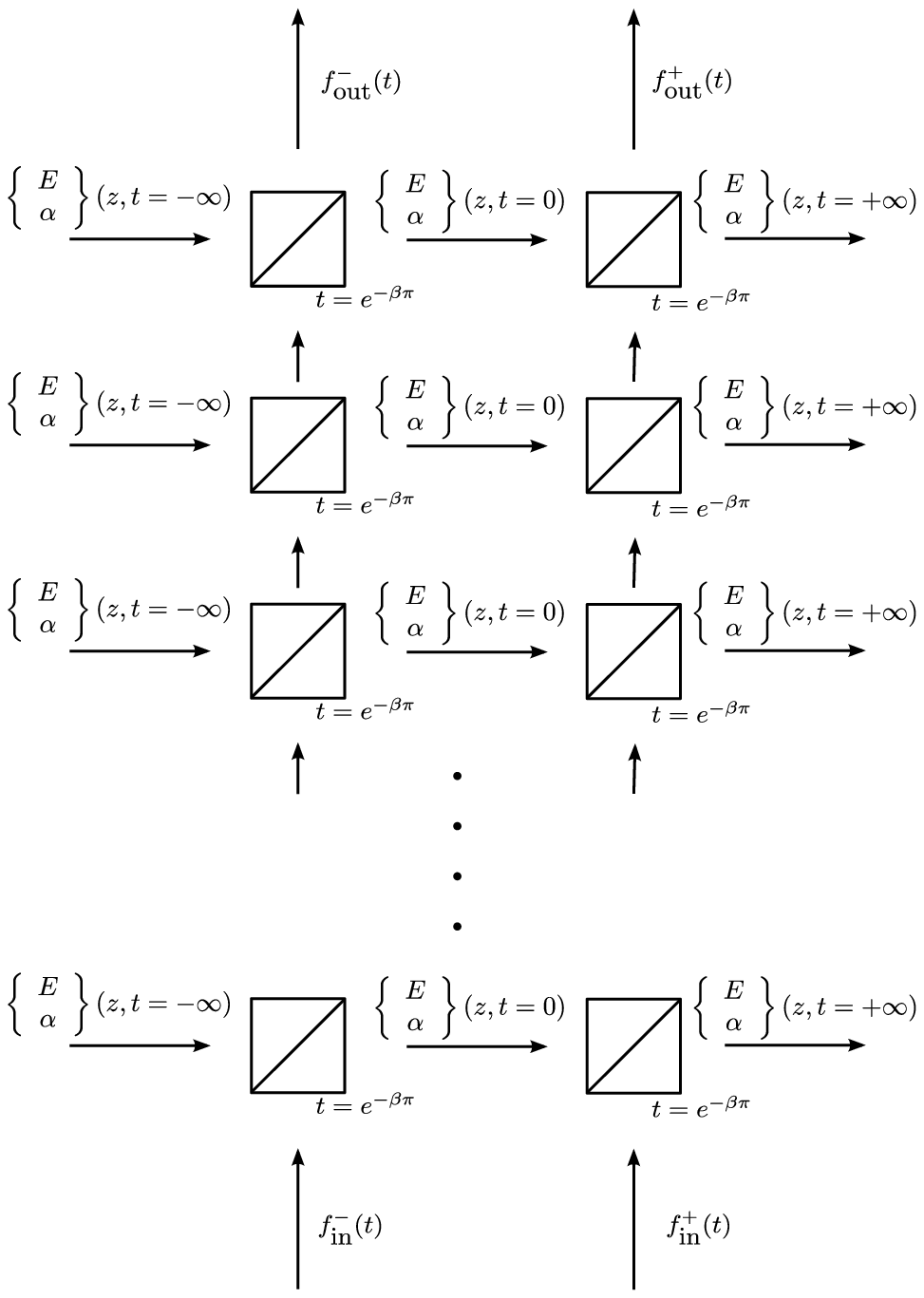}
\par\end{centering}

\caption{Network of analogue beamsplitters relevant to a transverse CRIB memory
modeled as a large number of optically gradient echo memories. \label{fig:beamsplitter2}}

\end{figure}

Initial theoretical treatments of the efficiency of two-level controlled
reversible inhomogeneous broadening (CRIB) echoes \cite{sang07} investigated
situations where the controlled detunings of the ions were not correlated
with the position the atom. Such transverse broadening would arise
from a microscopic broadening mechanism. This situation can be modeled
by a large number of optically thin gradient echo sub-memories in
series. The relevant network of analogue beamsplitters is shown in
Fig.~\ref{fig:beamsplitter2}. If there are $M$ sub memories, the
transmitted input pulse will be attenuated by $\exp(-2\beta\pi M)$.
There are $M$ paths a photon can take from the input pulse $f_{\text{in}}^{-}(t)$
to the output pulse $f_{\text{out}}^{+}(t)$, each of which involves
two reflections and $M-1$transmissions from our analogue beamsplitters.
Because the echo from each sub-memory combines in phase with that
from the previous memory these paths all combine constructively to
give an efficiency for the echo of $M^{2}(1-\exp(-2\beta\pi))^{2}\exp(-2\beta\pi(M-1))$.
Taking the limit where each sub-memory is optically thin ($\beta$
small) we arrive at an efficiency for the memory with microscopic
broadening

\[
\text{Efficiency}=4\beta^{2}\pi^{2}M^{2}\exp(-2\beta\pi M)=d^{2}e^{-d}\]

Here $d=2\beta\pi M$ is the optical depth, (the logarithm of the
ratio of the energies of the incident an transmitted pulses). This
result is the same as that derived in \cite{sang07} by solution of
the Maxwell-Bloch equations.

\begin{figure}
\begin{centering}
\includegraphics[width=1\columnwidth]{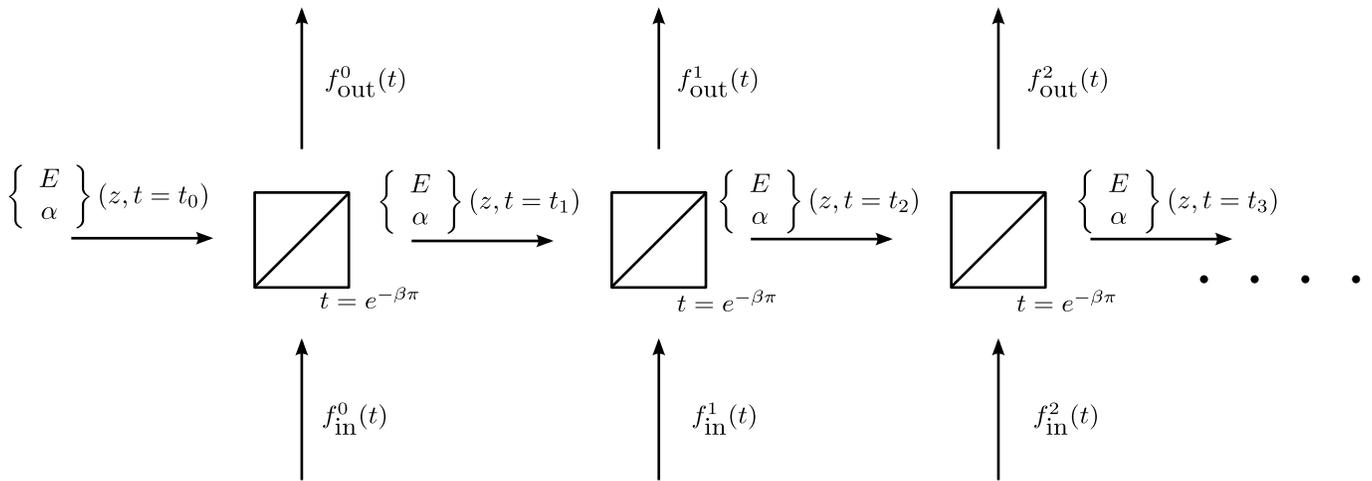}
\par\end{centering}

\caption{Network of analogue beamsplitters relevant to an optical gradient
echo memory where the field is switched a number of times\label{fig:beamsplitter3}}

\end{figure}

\begin{figure}
\begin{centering}
\includegraphics[width=0.9\columnwidth]{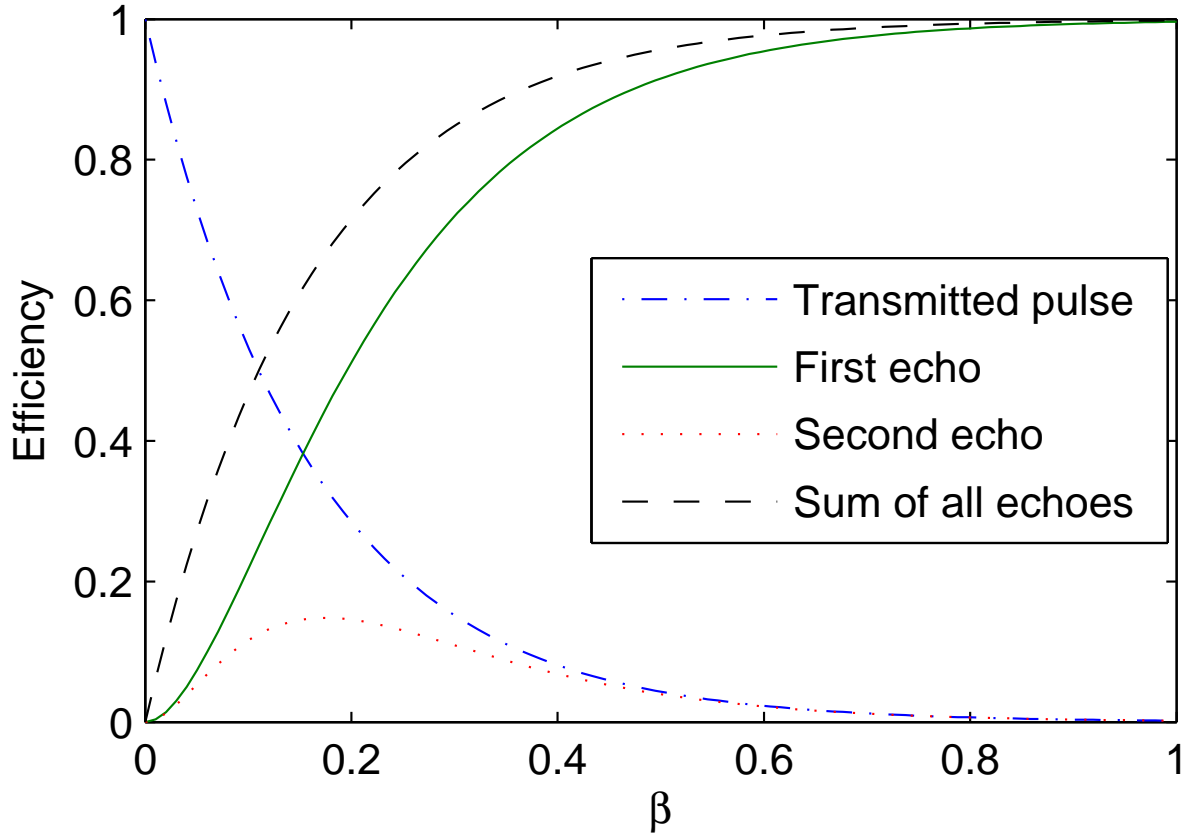}
\par\end{centering}

\caption{(Color online) Echo efficiency for an optical gradient echo as a function
of $\beta$ showing the effect of multiple switching. The ratio of
the energies of the input and transmitted pulse is given by $exp(-2\pi\beta)$.\label{fig:multi}}

\end{figure}

Another situation to which our beamsplitter analogy can be applied
is multiple switching of the broadening, both for longitudinal and
transverse broadenings. If the broadening is switched in polarity
after the first echo the excitation remaining in the sample will again
be rephased leading to another echo. Multiple echoes of the original
input pulse can be created in this way. The analogue beamsplitter
network for multiple switching in a gradient echo is shown in Fig.~\ref{fig:beamsplitter3},
from this network one can see that the the fraction of the input energy
coming out in the $k^{\text{th}}$echo is $(1-\exp(-2\beta\pi))^{2}\exp(-2\beta\pi(k-1))$.
From the beamsplitter network it can be seen that with multiple swicthing
the energy eventually leaves the sample in either as the transmitted
pulse or as one of the echoes. The efficiencies of the echos as a
function of $\beta$ are plotted in Fig.~\ref{fig:multi}.

The beam splitter analogy can be extended further to the case of multiple
switching from a system with microscopic broadening. The network of
analogue beam splitters in this situation is similar to Figs.~\ref{fig:beamsplitter2}
but with a 2D array of beam splitters. The amplitude of each of the
multiple echoes can be calculated by summing the amplitudes of all
the possible paths through the beamsplitter network. This leads to
the following expression for the portion of the incident energy that
is output as the $p{}^{\text{th}}$ echo

\[
e_{p}=t^{M+p}\left|\sum_{k=1}^{p}\left(\begin{array}{c}
p-1\\
k-1\end{array}\right)\frac{1}{k!}\left(-\frac{r^{2}}{t^{2}}\right)^{k}\right|^{2}\]

\begin{figure}
\begin{centering}
\includegraphics[width=0.9\columnwidth]{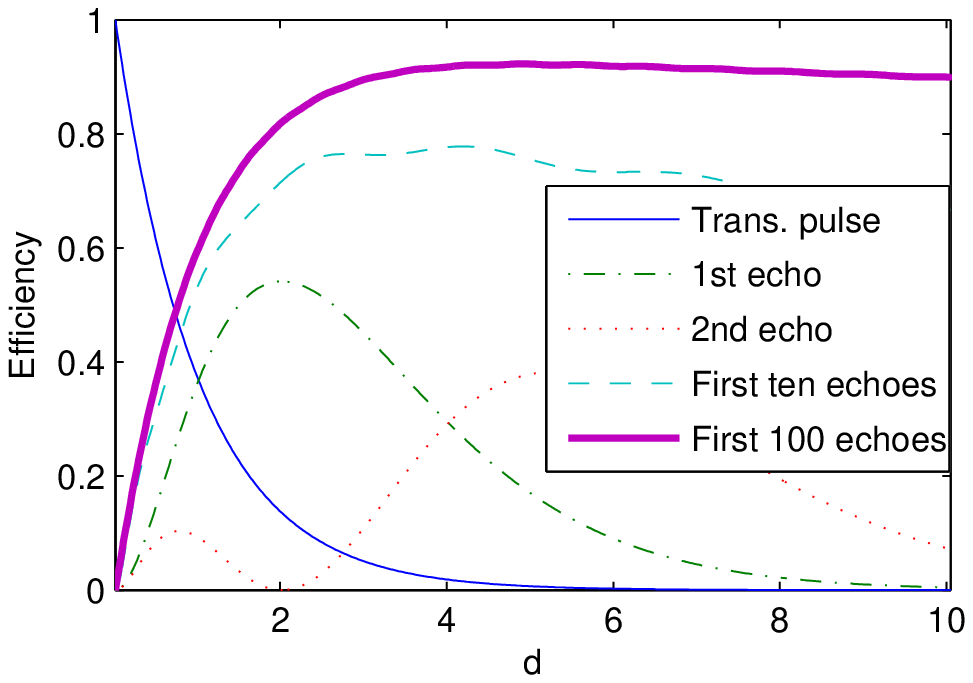}
\par\end{centering}

\caption{(Color online) Echo efficiency as a function of optical depth for
transverse broadening CRIB echo showing the effects of multiple switching.
As is shown in the figure efficiencies significantly greater than
54\% can be achieved with multiple switching but 100\% efficiency
is approached relatively slowly as the number of echoes increases.
Approximately 100 echoes are needed to achieve greater than 90\% efficiency.\label{fig:micromulti}}

\end{figure}

Here $r=(1-\exp(-2\beta\pi)$ and $t=\exp(-\beta\pi)$ are the amplitude
transmission and reflection coefficients of the analogue beamsplitters.
Taking the limit a large number of optically thin memories. We get
the following expression for the efficiency of the $p{}^{\text{th}}$
echo as\[
e_{p}=\exp(-d)\left|\sum_{k=1}^{p}\left(\begin{array}{c}
p-1\\
k-1\end{array}\right)\frac{1}{k!}(-d)^{k}\right|^{2}\]
where $d$ is the optical depth. As is shown in Fig.~\ref{fig:micromulti}
combined efficiencies significantly greater than 54\% can be achieved
with multiple switching. 

\section{Conclusion}

In conclusion we have presented an analytic treatment of the optical
gradient echo, or longitudinal CRIB echo. We have shown that it is
completely efficient in the case of large optical depth and that recall
efficiencies of 100\% can also be obtained for optically thin samples
by multiple switching. By modeling a system with transverse CRIB by
a large number of optically thin gradients can calculate echo efficiencies
in this case also. As has been shown elsewhere the maximum echo efficiency
is $4/e^{2}\approx54\%$. Multiple switching can improve this overall
efficiency with efficiencies of $>90\%$ possible for the sum of the
first 100 echoes.


\begin{thebibliography}{17}
\expandafter\ifx\csname natexlab\endcsname\relax\def\natexlab#1{#1}\fi
\expandafter\ifx\csname bibnamefont\endcsname\relax
  \def\bibnamefont#1{#1}\fi
\expandafter\ifx\csname bibfnamefont\endcsname\relax
  \def\bibfnamefont#1{#1}\fi
\expandafter\ifx\csname citenamefont\endcsname\relax
  \def\citenamefont#1{#1}\fi
\expandafter\ifx\csname url\endcsname\relax
  \def\url#1{\texttt{#1}}\fi
\expandafter\ifx\csname urlprefix\endcsname\relax\def\urlprefix{URL }\fi
\providecommand{\bibinfo}[2]{#2}
\providecommand{\eprint}[2][]{\url{#2}}

\bibitem[{\citenamefont{Knill et~al.}(2001)\citenamefont{Knill, Laflamme, and
  Milburn}}]{knil01}
\bibinfo{author}{\bibfnamefont{E.}~\bibnamefont{Knill}},
  \bibinfo{author}{\bibfnamefont{R.}~\bibnamefont{Laflamme}}, \bibnamefont{and}
  \bibinfo{author}{\bibfnamefont{G.~J.} \bibnamefont{Milburn}},
  \bibinfo{journal}{Nature} \textbf{\bibinfo{volume}{409}}, \bibinfo{pages}{46}
  (\bibinfo{year}{2001}).

\bibitem[{\citenamefont{Simon et~al.}(2007)\citenamefont{Simon, de~Riedmatten,
  Afzelius, Sangouard, Zbinden, and Gisin}}]{simo07}
\bibinfo{author}{\bibfnamefont{C.}~\bibnamefont{Simon}},
  \bibinfo{author}{\bibfnamefont{H.}~\bibnamefont{de~Riedmatten}},
  \bibinfo{author}{\bibfnamefont{M.}~\bibnamefont{Afzelius}},
  \bibinfo{author}{\bibfnamefont{N.}~\bibnamefont{Sangouard}},
  \bibinfo{author}{\bibfnamefont{H.}~\bibnamefont{Zbinden}}, \bibnamefont{and}
  \bibinfo{author}{\bibfnamefont{N.}~\bibnamefont{Gisin}},
  \bibinfo{journal}{Phys. Rev. Lett.} \textbf{\bibinfo{volume}{98}},
  \bibinfo{pages}{190503} (\bibinfo{year}{2007}).

\bibitem[{\citenamefont{Julsgaard et~al.}(2004)\citenamefont{Julsgaard,
  Sherson, Cirac, Fiurasek, and Polzik}}]{juls04}
\bibinfo{author}{\bibfnamefont{B.}~\bibnamefont{Julsgaard}},
  \bibinfo{author}{\bibfnamefont{J.}~\bibnamefont{Sherson}},
  \bibinfo{author}{\bibfnamefont{J.~I.} \bibnamefont{Cirac}},
  \bibinfo{author}{\bibfnamefont{J.}~\bibnamefont{Fiurasek}}, \bibnamefont{and}
  \bibinfo{author}{\bibfnamefont{E.~S.} \bibnamefont{Polzik}},
  \bibinfo{journal}{Nature} \textbf{\bibinfo{volume}{432}},
  \bibinfo{pages}{482} (\bibinfo{year}{2004}).

\bibitem[{\citenamefont{Fleischhauer and Lukin}(2002)}]{flei02}
\bibinfo{author}{\bibfnamefont{M.}~\bibnamefont{Fleischhauer}}
  \bibnamefont{and} \bibinfo{author}{\bibfnamefont{M.~D.} \bibnamefont{Lukin}},
  \bibinfo{journal}{Phys. Rev. A} \textbf{\bibinfo{volume}{65}},
  \bibinfo{pages}{022314} (\bibinfo{year}{2002}).

\bibitem[{\citenamefont{Gorshkov et~al.}(2007)\citenamefont{Gorshkov,
  Andr{\'e}, Lukin, and S{\o}rensen}}]{gors07}
\bibinfo{author}{\bibfnamefont{A.~V.} \bibnamefont{Gorshkov}},
  \bibinfo{author}{\bibfnamefont{A.}~\bibnamefont{Andr{\'e}}},
  \bibinfo{author}{\bibfnamefont{M.~D.} \bibnamefont{Lukin}}, \bibnamefont{and}
  \bibinfo{author}{\bibfnamefont{A.~S.} \bibnamefont{S{\o}rensen}},
  \bibinfo{journal}{Phys. Rev. A} \textbf{\bibinfo{volume}{76}},
  \bibinfo{pages}{033806} (\bibinfo{year}{2007}).

\bibitem[{\citenamefont{Moiseev and Kroll}(2001)}]{mois01}
\bibinfo{author}{\bibfnamefont{S.~A.} \bibnamefont{Moiseev}} \bibnamefont{and}
  \bibinfo{author}{\bibfnamefont{S.}~\bibnamefont{Kroll}},
  \bibinfo{journal}{Phys. Rev. Lett.} \textbf{\bibinfo{volume}{87}},
  \bibinfo{pages}{173601} (\bibinfo{year}{2001}).

\bibitem[{\citenamefont{Kraus et~al.}(2006)\citenamefont{Kraus, Tittel, Gisin,
  Nilsson, Kroll, and Cirac}}]{krau06}
\bibinfo{author}{\bibfnamefont{B.}~\bibnamefont{Kraus}},
  \bibinfo{author}{\bibfnamefont{W.}~\bibnamefont{Tittel}},
  \bibinfo{author}{\bibfnamefont{N.}~\bibnamefont{Gisin}},
  \bibinfo{author}{\bibfnamefont{M.}~\bibnamefont{Nilsson}},
  \bibinfo{author}{\bibfnamefont{S.}~\bibnamefont{Kroll}}, \bibnamefont{and}
  \bibinfo{author}{\bibfnamefont{J.~I.}~\bibnamefont{Cirac}},
  \bibinfo{journal}{Phys. Rev. A} \textbf{\bibinfo{volume}{73}},
  \bibinfo{pages}{020302(R)} (\bibinfo{year}{2006}),

\bibitem[{\citenamefont{Chanelire et~al.}(2005)\citenamefont{Chanelire,
  Matsukevich, Jenkins, Lan, Kennedy, and Kuzmich}}]{chan05}
\bibinfo{author}{\bibfnamefont{T.}~\bibnamefont{Chanelire}},
  \bibinfo{author}{\bibfnamefont{D.~N.} \bibnamefont{Matsukevich}},
  \bibinfo{author}{\bibfnamefont{S.~D.} \bibnamefont{Jenkins}},
  \bibinfo{author}{\bibfnamefont{S.-Y.} \bibnamefont{Lan}},
  \bibinfo{author}{\bibfnamefont{T.~A.~B.} \bibnamefont{Kennedy}},
  \bibnamefont{and} \bibinfo{author}{\bibfnamefont{A.}~\bibnamefont{Kuzmich}},
  \bibinfo{journal}{Nature} \textbf{\bibinfo{volume}{428}},
  \bibinfo{pages}{833} (\bibinfo{year}{2005}).

\bibitem[{\citenamefont{Eisaman et~al.}(2005)\citenamefont{Eisaman, Andr,
  Massou, Fleischhauer, Zibrov, and Lukin}}]{eisa05}
\bibinfo{author}{\bibfnamefont{M.~D.} \bibnamefont{Eisaman}},
  \bibinfo{author}{\bibfnamefont{A.}~\bibnamefont{Andr}},
  \bibinfo{author}{\bibfnamefont{F.}~\bibnamefont{Massou}},
  \bibinfo{author}{\bibfnamefont{M.}~\bibnamefont{Fleischhauer}},
  \bibinfo{author}{\bibfnamefont{A.~S.} \bibnamefont{Zibrov}},
  \bibnamefont{and} \bibinfo{author}{\bibfnamefont{M.~D.} \bibnamefont{Lukin}},
  \bibinfo{journal}{Nature} \textbf{\bibinfo{volume}{438}},
  \bibinfo{pages}{837} (\bibinfo{year}{2005}).

\bibitem[{\citenamefont{Appel et~al.}(2008)\citenamefont{Appel, Figueroa,
  Korystov, Lobino, and Lvovsky}}]{appe08}
\bibinfo{author}{\bibfnamefont{J.}~\bibnamefont{Appel}},
  \bibinfo{author}{\bibfnamefont{E.}~\bibnamefont{Figueroa}},
  \bibinfo{author}{\bibfnamefont{D.}~\bibnamefont{Korystov}},
  \bibinfo{author}{\bibfnamefont{M.}~\bibnamefont{Lobino}}, \bibnamefont{and}
  \bibinfo{author}{\bibfnamefont{A.~I.} \bibnamefont{Lvovsky}},
  \bibinfo{journal}{Phys. Rev. Lett.} \textbf{\bibinfo{volume}{100}},
  \bibinfo{pages}{093602} (\bibinfo{year}{2008}).

\bibitem[{\citenamefont{Sangouard et~al.}(2007)\citenamefont{Sangouard, Simon,
  Afzelius, and Gisin}}]{sang07}
\bibinfo{author}{\bibfnamefont{N.}~\bibnamefont{Sangouard}},
  \bibinfo{author}{\bibfnamefont{C.}~\bibnamefont{Simon}},
  \bibinfo{author}{\bibfnamefont{M.}~\bibnamefont{Afzelius}}, \bibnamefont{and}
  \bibinfo{author}{\bibfnamefont{N.}~\bibnamefont{Gisin}},
  \bibinfo{journal}{Phys. Rev. A} \textbf{\bibinfo{volume}{75}},
  \bibinfo{pages}{032327} (\bibinfo{year}{2007}).

\bibitem[{\citenamefont{Alexander et~al.}(2006)\citenamefont{Alexander,
  Longdell, Sellars, and Manson}}]{alex06}
\bibinfo{author}{\bibfnamefont{A.~L.} \bibnamefont{Alexander}},
  \bibinfo{author}{\bibfnamefont{J.~J.} \bibnamefont{Longdell}},
  \bibinfo{author}{\bibfnamefont{M.~J.} \bibnamefont{Sellars}},
  \bibnamefont{and} \bibinfo{author}{\bibfnamefont{N.~B.}
  \bibnamefont{Manson}}, \bibinfo{journal}{Phys. Rev. Lett.}
  \textbf{\bibinfo{volume}{96}}, \bibinfo{pages}{043602}
  (\bibinfo{year}{2006}).

\bibitem[{\citenamefont{Hetet et~al.}(2008)\citenamefont{Hetet, Longdell,
  Alexander, Lam, and Sellars}}]{hete08}
\bibinfo{author}{\bibfnamefont{G.}~\bibnamefont{Hetet}},
  \bibinfo{author}{\bibfnamefont{J.~J.} \bibnamefont{Longdell}},
  \bibinfo{author}{\bibfnamefont{A.~L.} \bibnamefont{Alexander}},
  \bibinfo{author}{\bibfnamefont{P.~K.} \bibnamefont{Lam}}, \bibnamefont{and}
  \bibinfo{author}{\bibfnamefont{M.~J.} \bibnamefont{Sellars}},
  \bibinfo{journal}{Phys. Rev. Lett.} \textbf{\bibinfo{volume}{100}},
  \bibinfo{pages}{023601} (\bibinfo{year}{2008}).

\bibitem[{\citenamefont{H{\'e}tet et~al.}(2008)\citenamefont{H{\'e}tet,
  Longdell, Sellars, Lam, and Buchler}}]{hete08b}
\bibinfo{author}{\bibfnamefont{G.}~\bibnamefont{H{\'e}tet}},
  \bibinfo{author}{\bibfnamefont{J.~J.} \bibnamefont{Longdell}},
  \bibinfo{author}{\bibfnamefont{M.~J.} \bibnamefont{Sellars}},
  \bibinfo{author}{\bibfnamefont{P.~K.} \bibnamefont{Lam}}, \bibnamefont{and}
  \bibinfo{author}{\bibfnamefont{B.~C.} \bibnamefont{Buchler}},
  \emph{\bibinfo{title}{Bandwidth and dynamics of the gradient echo memory}}
  (\bibinfo{year}{2008}), \eprint{arXiv:0801.3860}.

\bibitem[{\citenamefont{Alexander et~al.}(2007)\citenamefont{Alexander,
  Longdell, Sellars, and Manson}}]{alex07}
\bibinfo{author}{\bibfnamefont{A.~L.} \bibnamefont{Alexander}},
  \bibinfo{author}{\bibfnamefont{J.~J.} \bibnamefont{Longdell}},
\bibinfo{author}{\bibfnamefont{M.~J.} \bibnamefont{Sellars}},
  \bibnamefont{and} \bibinfo{author}{\bibfnamefont{N.~B.}
  \bibnamefont{Manson}}, \bibinfo{journal}{J. Lumin}
  \textbf{\bibinfo{volume}{127}}, \bibinfo{pages}{94} (\bibinfo{year}{2007}).

\bibitem[{\citenamefont{Crisp}(1970)}]{cris70}
\bibinfo{author}{\bibfnamefont{M.~D.} \bibnamefont{Crisp}},
  \bibinfo{journal}{Phys. Rev. A} \textbf{\bibinfo{volume}{1}},
  \bibinfo{pages}{1604} (\bibinfo{year}{1970}).

\bibitem[{\citenamefont{Fleischhauer and Lukin}(2000)}]{flei00}
\bibinfo{author}{\bibfnamefont{M.}~\bibnamefont{Fleischhauer}}
  \bibnamefont{and} \bibinfo{author}{\bibfnamefont{M.~D.} \bibnamefont{Lukin}},
  \bibinfo{journal}{Phys. Rev. Lett.} \textbf{\bibinfo{volume}{84}},
  \bibinfo{pages}{5094} (\bibinfo{year}{2000}).

\end{thebibliography}
\end{document}